\documentclass[prl,twocolumn,showpacs,floatfix]{revtex4}
\usepackage{graphics,epsfig,amsfonts,amssymb,amsmath}
\begin{document}
\title{Absence of scaling in transport through two-dimensional
   nanoparticle arrays}
\author{V. Est\'evez}
\author{E. Bascones}
\affiliation{Instituto de Ciencia de Materiales de Madrid,
ICMM-CSIC, Cantoblanco, E-28049 Madrid (Spain).}
\email{vestevez@icmm.csic.es,leni@icmm.csic.es}

\begin{abstract}
We analyze the transport in disordered two-dimensional 
nanoparticle arrays. We show that 
the commonly used scaling hypothesis to fit the I-V curves 
does not describe the electronic transport in these systems.
On the contrary, close to the threshold voltage $V_T$  
the current depends linearly on 
$(V-V_T)$. 
This linear behavior 
is observed for at least five decades in $(V-V_T)$.  
Fitting the I-V curves at larger voltages
to a scaling power-law
$I \propto (V/V_T-1)^\xi$ results in fitting parameters which depend on the
range of voltages used and in wrong values for $V_T$.  
Our results urge to change the picture of electronic transport in 
disordered nanoparticle arrays used in the last two decades. 

\end{abstract}
%\pacs{75.10.Jm, 75.10.Lp, 75.30.Ds}
\maketitle

Since the pioneering work of Middleton and Wingreen\cite{middletonwingreen93}  (MW) in 1993 the 
transport in disordered nanoparticle arrays  has been interpreted in terms of 
a threshold voltage $V_T$, the minimum bias voltage necessary to allow the flow 
of current, and I-V curves with power-law behavior $ I \sim (V/V_T-1)^\xi$ .  
For charge disordered arrays and short-range interactions MW predicted 
$\xi=1$ 
and $5/3$ for one (1D) and two dimensions (2D) respectively. This power law is 
supposed to hold above, but arbitrarily close to the threshold voltage.

In the last decade a wide variety of 2D arrays has been available 
and their transport properties studied\cite{grzelczak10,zabet08}. Experimental 
I-V characteristics have 
been systematically discussed in terms of these 
power-laws\cite{rimberg95,black00,jaeger01,ancona01,ancona03,romerodrndic05,elteto1d05,blunt07,tan09,sachser09}.  The exponent 
observed is, on the other hand, larger, $\xi \geq 2$ than expected in most of the
experiments and frequently sample-dependent.  
Deviations of the observed exponent from the predicted value 
have been often interpreted in terms of a dimensionality of the experimental 
set-up larger than two.  But the scaling exponent found in quasi-one 
dimensional strips was $\xi \sim 2$.  
To claim scaling behavior at least two decades in the scaling parameter should 
be desirable.  Experimentally the  power laws have never been 
observed in such a large range of voltages but, they have been in most cases 
restricted to less than a decade, somewhere in the region 
$(V/V_T-1) \sim  0.04-10$.

Numerical  confirmation of the 2D exponent $\xi= 5/3$ has been 
also elusive. For short range interactions, MW found $\xi \sim 2.0$ for 
$(V/V_T-1) \sim  0.1$. Later, Jha and Middleton\cite{jhamiddleton05} 
failed  to define a proper power-law.    
Similarly, for long-range interactions\cite{kaplan03}
$\xi \sim 2.0$. The
discrepancy was attributed to finite-size effects\cite{middletonwingreen93,jhamiddleton05}. Interestingly, Jha and
Middleton\cite{jhamiddleton05} argued that the voltages at which the exponents 
$\sim 2$ are found correspond 
to a region outside the putative MW regime.      
In this paper we show that the reason for all these discrepancies is that 
the power-law scaling description of MW  fails. 

The prediction of MW is based on the assumption that close to threshold 
the current flows through $N_{ch}$ independent channels, each of them 
driving a current linearly dependent on  $(V-V_T)$. The number of channels 
depends on the dimensionality. For 
1D systems $N_{ch}=1$. On the basis of a mapping of 
the  current flow to a model of interface 
growth\cite{KPZ}, which neglects the role of the contact junctions in
determining the current, 
they concluded that in 2D systems 
$N_{ch}\sim (V/V_T-1)^{2/3}$, which together with 
the linear current of a 1D channel gives 
$\xi= 5/3$.

Recently  we confirmed that 1D arrays show a linear 
dependence close 
to threshold\cite{nosotrosprb08}. This linearity lasts for at least 
five orders in 
magnitude, see Fig. 1, but  it disappears at voltages much smaller  than 
those at which both experiments and previous numerical calculations were 
performed .  Linearity arises from the voltage dependence of the tunneling 
rate at the contact junction (between  array and  electrodes) which acts 
as a bottle-neck for the current. 
Having in mind the influence of the contact junctions in 1D arrays close to
threshold, 
we expect that in 2D systems the current is carried by a single channel and 
linear I-V curves with slope determined by the resistance of the contact 
junctions and the failure of MW prediction. At the voltages at which 
new conduction channels 
open the linear dependence of the first channel has disappeared
invalidating MW assumptions. 

In order  to test the validity of MW scaling argument we have carried 
systematic numerical simulations in 2D systems. We have found 
that, as in 1D, in charge disordered 2D arrays close to threshold the current 
is carried by a single channel, and depends linearly on voltage. This dependence
lasts for several 
decades, but disappears at small voltages, not accessible experimentally.
With increasing voltage the I-V curves show a crossover which in large arrays
resemble a super-linear power-law.
Previous claims of scaling and power-laws have 
been done in the range of voltages at which we observe this crossover.  
However, fitting the crossover to a power-law produces threshold voltages 
$V_T$ and exponents $\xi$ which depend on the range of voltages used in the 
fitting and fail to give the correct value of the threshold.

We consider an array of $m \times n$ metallic islands in the classical Coulomb
blockade regime, $\delta \ll  K_BT
\ll E_c$ with $E_c$ the  charging energy, $\delta$ the single particle level
spacing, $T$  the temperature and $K_B$ the Boltzmann constant. 
Temperature is then taken equal to zero. 
The array is placed in between two electrodes at voltages $\pm V/2$. The 
islands are separated
between themselves and from the contacts by tunnel junctions. 
Except otherwise indicated we 
assume all the junctions to have the same resistance.
Electronic 
interactions are assumed to be finite only when the charges are in the same 
conductor, i.e. capacitive coupling between different conductors vanishes.  
The electronic charge is taken equal to unity. 
Transport is treated at the sequential tunneling 
level. 
To compute the current we use a Monte-Carlo simulation,
described previously\cite{nosotrosprb08,likharev89}. We have studied clean and 
disordered arrays with square and triangular lattices and 
three types of disorder: charge
disorder, resistance disorder and structural disorder, i.e. voids in the 
lattice.

Before discussing two-dimensional systems we review the transport in  1D 
charge disordered $N$-particle arrays \cite{nosotrosprb08} and show that while linearity 
lasts for several orders of magnitude, proper scaling in the MW sense is not 
present. As discussed above, the current is blocked up to a threshold 
voltage $V_T$ which depends on the disorder configuration. Below $V_T$ charges 
entering from the electrodes pile-up inside the array and create charge 
gradients which overcome the upward steps in the disorder 
potential\cite{middletonwingreen93}.   At $V_T$ a charge entering from the 
electrodes is able to flow through all the array. 
Above, but very close to the threshold the entrance of charges onto the array 
act as a bottle neck for the current. The current can be approximated by the 
tunneling rate at the contact junction which controls the  entrance of 
charges to the array\cite{nosotrosprb08}. 
This rate increases linearly with $(V-V_T)$ resulting in a current
$I=R_{bn}^{-1}(V-V_T)$.
 Here $R_{bn}$ is the resistance of the bottle neck junction. 
To understand this equation it is important to take into account the way 
in which the voltage  (not to be confused with the total potential) drops
through the array.
For short-range interactions the voltage drops only at the contact 
junctions, between array and  electrodes\cite{nosotrosprb08}.

The linearity close to threshold lasts for several orders of magnitudes, see Fig. 1(a).   
The slope is independent of the array size, while the threshold voltage, when 
averaged over disorder configurations is proportional to the number of 
particles $<V_T> \approx N E_c$\cite{middletonwingreen93,nosotrosprb08}. This means that proper scaling of the 
current in terms of $(V/V_T-1)$ does not occur, as observed in 
Fig. 1 (b). Disagreement with MW  originates in the voltage drop 
through the array which they thought to be homogeneous. Deviations from linearity 
happen at small $(V-V_T)$ when the contact junction stops being the 
bottle-neck.

\begin{figure}
\leavevmode
\includegraphics[clip,width=0.5\textwidth]{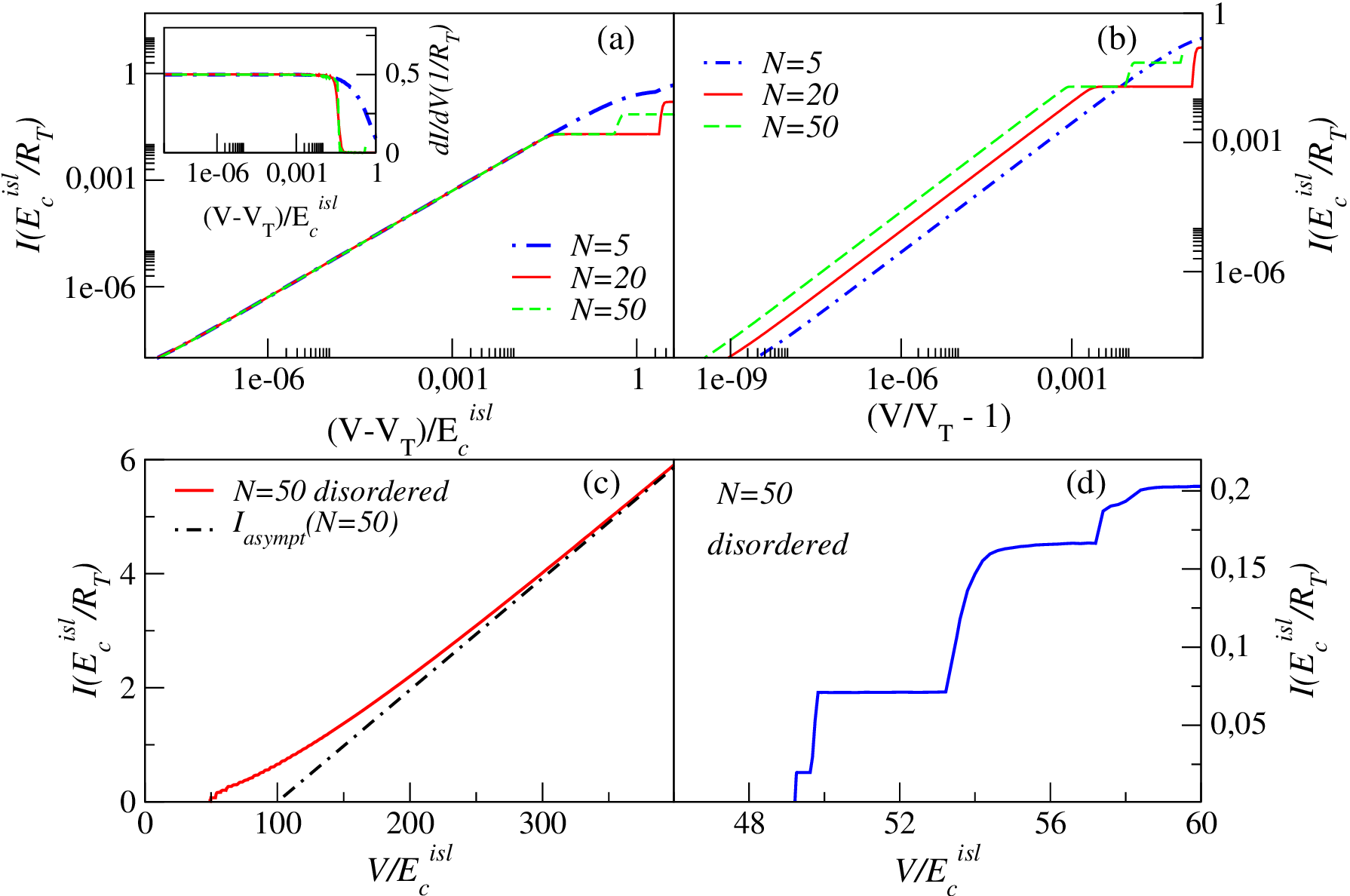}
\caption{(a) Main figure: I-V curves in logarithmic scale of 1D disordered arrays of 
different sizes. Inset: Derivative of the I-V curves in main figure. (b) Same as in (a) but with $(V/V_T-1)$ in the X-axis. (c) I-V curve of a 1D N=50 array in a large voltage regime compared with 
the high-voltage asymptotic I-V curve, see text. Same as in (c) but at smaller voltages where the Coulomb staircase can be clearly seen.}
\end{figure}

At high voltages ($V> 7 N E_c$)  the current approaches 
$1/R_{sum}(V-V_{offset})$ with  
$V_{offset} \approx 2 N E_C$, and $R_{sum}$   the sum of 
the tunnel resistances in series, see  Fig. 1(c).  In between these 
two linear regimes the current increases showing Coulomb staircase plateaux, 
see Fig. 1(d). Plateau-like behavior appears when the current is controlled 
by the tunneling 
through a junction which tunneling rate does not depend on the bias voltage\cite{nosotrosprb08}.

In 2D disordered arrays at voltages just above the threshold the current is 
carried by a single path. This is the path which requires the smallest pile-up 
of charges to overcome the disorder potential. Until a second path opens one 
might expect that the current looks like the one of a one-dimensional 
channel with slope controlled by the contact junction through which the 
charges enter. This is confirmed in Fig. 2. The linear behavior is observed in 
clean and disordered arrays with square or triangular lattice and in the 
presence of voids. It lasts for at least five orders in magnitude and
disappears at values of $(V-V_T)$ similar to those found in 1D arrays, 
much smaller than those in experiments and previous numerical simulations. 
The derivative of the I-V curve in disordered arrays, equal in 1D and 2D, 
confirms that a single channel drives the current, see 
inset in Fig.2(a). In contrast, in clean arrays 
$N_{ch} \propto N$ channels open at $V_T$.  

\begin{figure}
\leavevmode
\includegraphics[clip,width=0.5\textwidth]{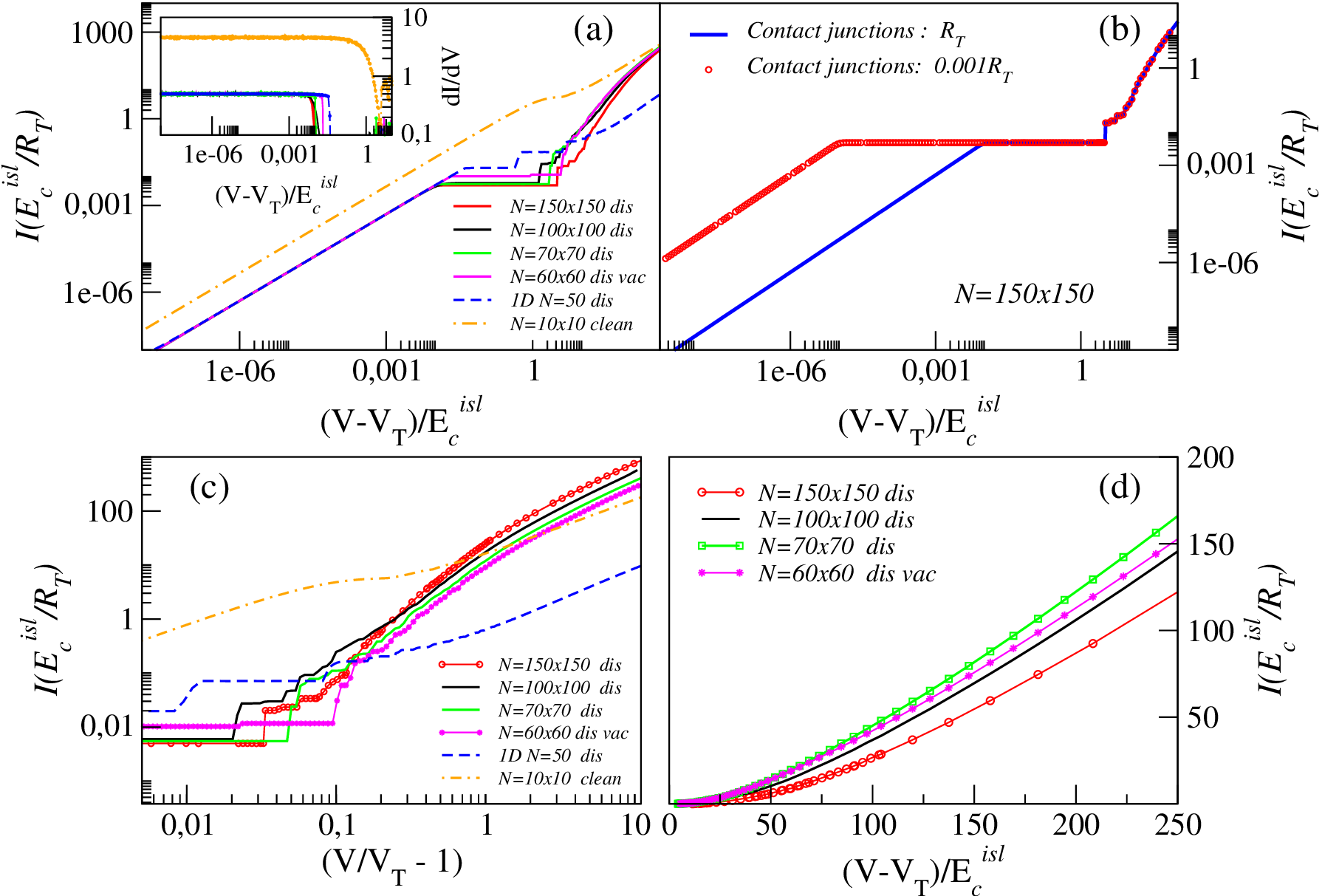}
\caption{(a) I-V curves at low voltages 
in logarithmic scale corresponding to  2D disordered
arrays of different sizes, a 50 islands disordered 1D array and a clean 10x10
square-lattice 2D clean array. The $60 \times 60$ square
lattice has structural disorder (vacancies) besides the charge disorder. 
Inset: derivatives of the I-V curves plotted in the main figure. 
(b) I-V curves for two $150 \times 150$ square lattice arrays with the same 
charge disorder configuration. In one of them the tunnel resistance of the 
contact junctions is 1000 times smaller.  
(c) Same as in main figure in (a) but in a different
voltage regime and with the axis scaled differently. (d) Same as in (c) for
some selected curves in normal scale with the same axis as in (a). 
}
\end{figure}

As in 1D, a linear dependence on $(V-V_T)$ does not mean scaling on 
$(V-V_T)/V_T$. This is seen in Fig. 2 (a) where all the curves, 
corresponding to  arrays with different $V_T$, 
show the same current in the linear regime 
in units of $V-V_T$. The whole lattice determines 
$V_T$, but a single contact junction, controls the 
slope of the current close to threshold. To emphasize this, in 
Fig.2 (b) we plot the I-V of a 
disordered array with all the resistances equal and the 
I-V of the same array but with contact resistances between electrodes 
and array one thousand times smaller than those between the islands. In the
array with small contact resistances the current is three orders of magnitude
larger at low voltages. This confirms that the contact junctions, and not the
lattice as usually assumed, control the
current close to threshold.  With increasing voltage it is the lattice who 
controls the current and the influence of the contact junction decreases.  

The scaling behavior discussed by MW was partly based on how new 
channels open to current flow.  Clearly, between the low voltage linear regime 
controlled by a single channel and the high voltage linear regime to which 
many channels contribute, there should be at least a crossover regime with 
channel opening. In Fig. 2 it is seen that in this crossover 
the I-V curves show clear steps 
with horizontal plateaux. Plateau-like features indicate that one or several
inner junctions, with tunneling rates independent of bias voltage, act as 
bottle-neck for the 
current. Similar plateaux where observed in \cite{rimberg95}. 
Steps are associated to channel opening.  The steps smooth with 
increasing array size as new channels open in smaller voltage intervals. 
On average, the current increases faster than linear and resembles a power-law 
in large arrays, see Fig. 2(d). As seen in Figs. 2(c) and 2(d), in the
crossover range of voltages, the current of an $N \times N$
disordered array does not scale with 
$ (V/V_T-1)$, nor $(V-V_T)/E_c$.
We note here that, as early discussed by Jha and
Middleton\cite{jhamiddleton05} the range of voltages where the superlinear
behavior is found $V-V_T > E_c$ is out of the {\it close to threshold} regime
discussed by MW. In fact, this range is closer to the high-voltage regime,
discussed below than to the {\it close to threshold} low voltage regime.   

To make connection with experimental results we have checked the 
fitting parameters which are obtained when the I-V curves in this 
regime are fitted to a scaling power law of the kind proposed 
by MW, $I = A (V/V_T-1)^\xi$. 
In Fig. 3 (a) and (b) we perform fittings to 
the I-V curve of a charge disordered $150 \times 150$ array, 
using this power-law expression, the value of
$V_T$, known theoretically, and the range of voltages 
plotted in each figure.  The values of the exponents that we obtain are 
similar
to the ones discussed in the literature. However, the fitting parameters, 
and in particular 
the scaling exponent, change considerably depending on how large it is 
the range of voltages used in the fitting, even if this range of voltages is 
quite small. 
This fact suggests that a power-law does not describe the current-voltage 
dependence. 

\begin{figure}
\leavevmode
\includegraphics[clip,width=0.5\textwidth]{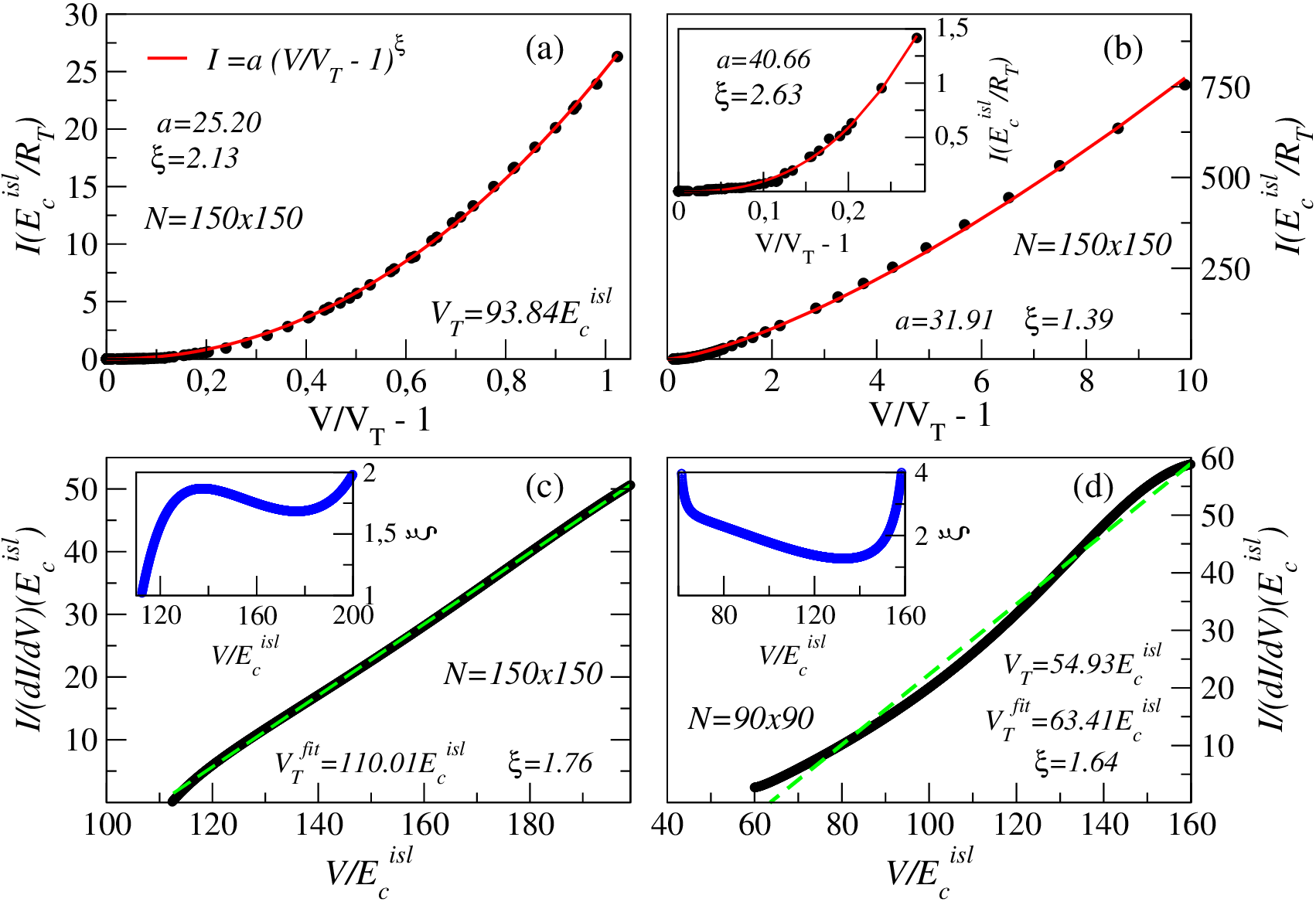}
\caption{(a) and (b) Main figures and inset: I-V curves for the same square
  lattice charge disordered $150 \times 150$ array with fittings
  to $I=A(V/V_T-1)^\xi$ within the range of voltages plotted, using the 
  theoretical  $V_T$, given in (a). 
  (c) and (d) $I/(dI/dV)$ for the same 
  array in (a) and (b) and for a $90 \times 90$ disordered array. To avoid the 
  noise in the derivatives, 
  the I-V curves had been previously fitted to a high order polynomial.   
  The smallest voltages are not shown to avoid spurious
  dependences induced by the polynomial fitting.
  Also shown the fittings to $\xi^{-1}(V-V_{T}^{fit})$, see text.
  Insets: Values of $\xi$ extracted from the 
   derivatives of (c) and (d) according to: $\xi^{-1}=d(I/(dI/dV))/dV$, see
   text. 
  }
\end{figure}

If such a power-law like 
were a good approximation to the 
current, plotting $I/(dI/dV)=\xi^{-1}(V-V_T)$ one could determine $\xi$ and 
$V_T$. This method 
has been used experimentally to extract these 
parameters\cite{ancona01,blunt07,ruffino07,sachser09,tan09}. In Fig. 3 (c) and (d) we show these functions 
for the $150 \times 150$  array in (a) and (b) and for a $90 \times 90$ array, 
with their corresponding fitting parameters. Notice that the $V_T^{fit}$
obtained in the fitting does not equal the true one. 
One can go even further and derive these curves. Such a derivative should 
give a constant $\xi^{-1}$. As shown in the insets of Figs. 3 (c) and (d)
these derivatives while giving $\xi \sim 2$  are far from 
being constant.  This fact confirms that the crossover is neither described by a power-law function and the failure of 
MW description in 2D arrays.

One might ask how important are finite-size effects  and if  the crossover 
region could extend to smaller voltages and converge to the predicted 
power-law in much larger systems. 
No features in our data suggests this to be the case. Within the range of 
sizes analyzed, with the number 
of particles
varying between 400 and 29000,  
we have not found any 
systematic dependence of the voltage at which the superlinear 
crossover starts as a function of array size
One could think that our lattices are still small. 
We now argue that this is not the case.

Let us consider an $N \times N$ square lattice. For larger N, on average, 
the new channels could open for smaller values of $(V-V_T)$  as increasing  
the number of rows increases the possibilities to find a new path with a small 
threshold. On the other hand as the number of columns becomes larger the 
contact junction of the early open channels stops being the bottle-neck for 
smaller voltages and their linear current-voltage contribution  is substituted 
by a plateau. Thus, it does not seem possible to satisfy the two assumptions of 
Middleton and Wingreen (channel opening and linear dependence) 
at the same time. But even if there were a way in which 
both assumptions were satisfied the linear behavior of each path would refer 
to  its own path threshold voltage $I \sim (V-V_{T,path})$. $V_{T,path}$ is 
larger than the 
array $V_T$, which is the  voltage of the {\it first} 
path which opens. This means that this linearity would not scale 
with $(V/V_T-1)$ as assumed by MW implying that the derived equations 
would not be correct. 

We end with a brief discussion of the large voltage regime. 
At high voltages, $V > 8 N Ec $, and in the absence of voids, 
as in 1D
 the current is linear but extrapolate to zero at a finite offset 
voltage $V_{offset}$ which keeps memory of the interaction 
effects\cite{nosotrosprb08, kaplan03}.  
In the case of a square lattice, with  no voids nor resistance disorder, 
the high-voltage regime can be approximated by
$I=\sum_{i=rows} (V-V_{offset})/R_{sum,i}$ 
with $R_{sum,i}$ the sum of the junction resistances in row $i$ in series. 
This equation is valid for both clean and charge 
disordered systems, see Fig. 4 (a). 
On the other hand when the junction resistances are not all equal,
the current cannot be approximated by this expression, 
see Fig. 4 (b). This originates in the meandering of charges to avoid 
large resistance junctions. 
A similar effect appears in lattices with voids.

\begin{figure}
\leavevmode
\includegraphics[clip,width=0.42\textwidth]{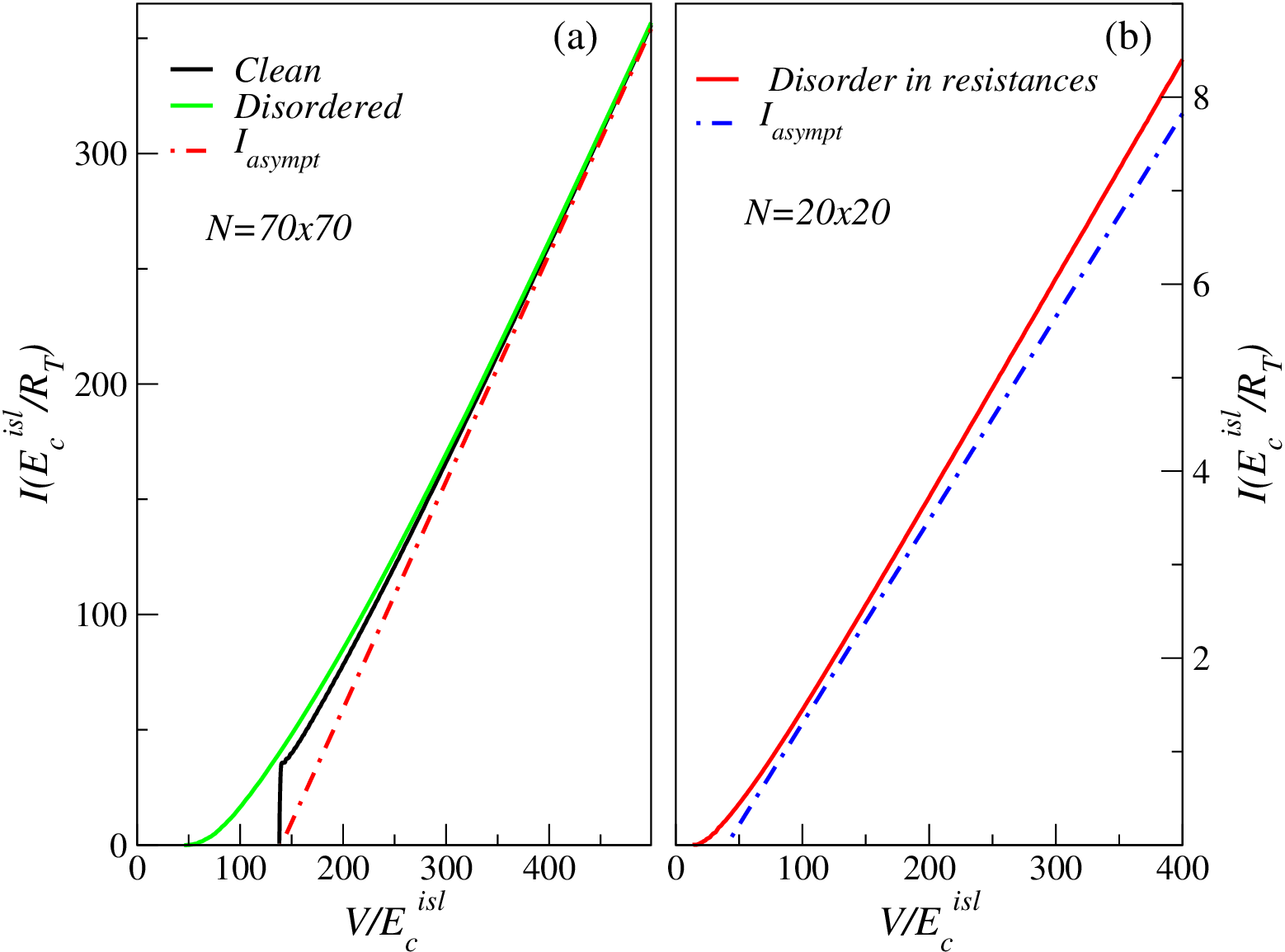}
\caption{(a) I-V curves for $70 \times 70$ square lattice arrays corresponding
  to clean and charge disordered 2D systems. 
  (b) I-V curve for a charge and resistance disordered $20
  \times 20$ square lattice. Resistances vary between $20 R_T$ and $84
  R_T$. The I-V is compared with the asymptotic I-V curve derived assuming 
 a potential drop equal to the one corresponding to adding the resistances in
 series, as discussed in the text. 
}
\end{figure}

In conclusion, we have shown that the scaling law of Middleton and 
Wingreen\cite{middletonwingreen93} and widely used since
then, does not describe the I-V of 2D disordered
arrays close to threshold. We find that close to threshold the current is
controlled by the contact junctions which act as bottle-neck and not by the 
whole lattice as it was assumed in that work. Contrary to the prediction of 
a power-law  with exponent $5/3$, close to threshold the current 
depends linearly on $(V-V_T)$.  In the crossover region at larger voltages, 
our calculations agree with experimental results when trying to fit the I-V 
curves to a power-law scaling curve. 
However, such fitting
results in meaningless fitting-parameters which depend on
the range of voltage considered and in wrong values for $V_T$.  
Our calculations urge to leave the scaling description for the 
transport in
2D systems, used during the last two decades.

Funding from Ministerio de Ciencia e Innovaci\'on through
Grants No. FIS2008-00124, FPI fellowship and Ram\'on y Cajal contract, and
from Consejer\'ia de Educaci\'on de la Comunidad Aut\'onoma de Madrid and CSIC
through Grants No. CCG07-CSIC/ESP-2323, CCG08-CSIC/ESP3518, PIE-200960I033 is 
acknowledged.

\end{document}